\documentclass[12pt,a4paper]{article}

\usepackage{amsmath,amssymb,amsfonts}
\usepackage{graphicx}
\usepackage[colorlinks,hypertexnames=false]{hyperref}
\usepackage{cite}

\begin{document}
\title{Magnetic field in vacuum of quantum spinor matter induced by a cosmic string in three-dimensional space}
\author{Yu.I.~Pylypchuk${}^1$,
P.O.~Nakaznyi${}^1$,\\
O.V.~Barabash${}^2$,
A.O.~Zaporozhchenko${}^2$,
V.M.~Gorkavenko${}^{2,3}$\vspace{0.5em}
\\
${}^1$ \it \small Institute of Physics and Technology, Igor Sikorsky Kyiv Polytechnic Institute,\\
\it \small 37, prospect Beresteiskyi, Kyiv 03056, Ukraine\\
${}^2$ \it \small Faculty of Physics, Taras Shevchenko National University of Kyiv, Ukraine,\\
\it \small 64, Volodymyrs'ka str., Kyiv 01601, Ukraine\\
${}^3$ \it \small Bogolyubov Institute for Theoretical Physics, National Academy of Sciences of Ukraine\\
\it \small 14-b, Metrolohichna str., Kyiv 03143, Ukraine}

\date{}

\maketitle
\setcounter{equation}{0}
\setcounter{page}{1}%

\begin{abstract}
    A linear magnetic topological defect (cosmic string) is modeled as a magnetic flux-carrying tube that is impenetrable to external spinor matter. The matter field is quantized in the background of this tube, with the most general set of boundary conditions ensuring both the tube’s impenetrability and the self-adjointness of the Dirac Hamiltonian operator. We compute the induced vacuum magnetic flux along the tube in $(3+1)$-dimensional space-time. It was shown that the requirement for the total induced vacuum magnetic flux to be finite restricts the admissible boundary conditions to only one choice: the MIT quark bag boundary condition. The dependence of the effect on the transverse size of the tube and the flux inside the tube was also analyzed.

    
\end{abstract}

\section{Introduction}

Various extensions of the Standard Cosmological Model predict the emergence of topological defects during the thermal evolution of the early Universe,  see, e.g., \cite{Jean,Buchmüller2023,Sar,Jones,Copeland,Copeland2010}. In particular, successive phase transitions associated with spontaneous symmetry breaking could have led to the formation of cosmic strings, monopoles, domain walls and textures, depending on the underlying symmetry-breaking pattern. In this paper, we focus on linear topological defects, also known as cosmic strings, which may persist today, see, e.g., the reviews \cite{Vil3,Ki3}. Such objects can reveal themselves through distinct astrophysical phenomena, such as the generation of gravitational waves \cite{Dam}, gamma-ray bursts \cite{Ber}, and ultra–high-energy cosmic rays \cite{Bhat}.

Inside the cosmic string, the field configuration of the phase with unbroken symmetry has energy and is characterized by mass per unit length of the string (or string tension) $\mu \sim \eta^2$, where $\eta$ is the energy scale of symmetry breaking ($\eta \sim m_H$, $m_H$ is the Higgs field mass). 
The energy (mass) concentrated in the core of a topological defect makes it a gravitational source. As a result, the exterior space-time acquires a conical geometry with a deficit angle $8\pi G \mu$: the squared length element in the outer region of the cosmic string is
\begin{equation}\label{1.2}
ds^2= dr^2+(1-4G \mu)^2 r^2 d\varphi^2+dz^2,
\end{equation}
where $G$ is Newton’s gravitational constant, and cylindrical coordinates are chosen with the symmetry axis aligned with the string. Model-independent constraints on the tension of the cosmic string from gravitational lensing are $G \mu \lesssim 10^{-6}$ \cite{Morganson,Christiansen}. Stronger constraints come from the analysis of the CMB \cite{Charnock} and gravitational waves \cite{Abbott}. Thus, the existing observational constraints rule out only cosmic strings that could have formed at extremely high energy scales, slightly below the scale of Grand Unified Theories.

In the Abelian Higgs model \cite{nielsen}, cosmic strings are formed during the spontaneous breaking of a gauge symmetry. The associated gauge field gives rise to a "magnetic" field that is confined to the string core. In the early Universe, this field need not be the familiar electromagnetic one; instead, it may correspond to an additional $U_X(1)$ gauge group beyond the Standard Model. It should be noted that such strings acquire an additional global characteristic — a quantized magnetic flux,
\begin{equation}\label{quantflux}
\Phi = \int\limits_{\rm core} d\sigma\, \mbox{\boldmath $\partial$}\times \textbf{A}=\oint d\mathbf{x} \cdot \mathbf{A} = \frac{2\pi n}{e_H},
\end{equation}
where $n \in \mathbb{Z}$ is an integer number (winding number), and $e_H$ is the coupling between the scalar (Higgs) condensate, which forms the string through spontaneous symmetry breaking, and the $U_X(1)$ gauge field.

For the quantized matter field outside the cosmic string interacting with the gauge field of the group $U_X(1)$ through the coupling $\tilde e$, the field solutions will be determined by the parameter ${e}\Phi/(2\pi)=n {\tilde e}/e_H$, which can also take non-integer values. We will use this important parameter in what follows. If, in addition to being charged under the $U_X(1)$ group, the matter fields also carry charge with respect to the electromagnetic group $U_{EM}(1)$, then the resulting induced current in the vacuum of electrically charged particles generates the usual magnetic field around the cosmic string, whose effects could, in principle, be probed observationally.

The study of the effects of spinor field vacuum polarisation in the presence of linear topological defects began in the 1990s. The first investigations were performed in the approximation of zero transverse size of such objects \cite{Gornicki,Flekkoy,Parwani,Sit96,Sit99,Bellucci14}. In further studies, vacuum effects were investigated, taking into account the internal structure of the cosmic string in certain theoretical models \cite{Bezerra10, Bezerra11, Bezerra12, Sousa16, Sousa17}. 

In this paper, we will assume that the internal structure of the cosmic string is unknown, as it is not precisely known how the cosmic string was formed. So, we will consider the general case, when the cosmic string is modeled as impenetrable for the matter field tube of finite radius $r_0\sim m_H^{-1}$ with "magnetic" field inside.  The complicated internal structure of the string will be revealed by selecting the appropriate boundary condition at the tube surface. It should also be noted that, although the "magnetic" field vanishes outside the string core, the gauge potential itself remains nonzero. This underlies the well-known Aharonov–Bohm effect \cite{Aha}.

In the case of a scalar matter field $\phi$, the boundary conditions on the tube can be of the Dirichlet type ($\phi|_{r=r_0}=0$), the Neumann type $(\partial_r \phi|_{r=r_0}=0)$, or (their combination) of the Robin type $(\cos \theta \,\phi + \sin \theta \, r \partial_r \phi)|_{r_0} =0$. A detailed analysis of the induced vacuum energy in these cases was carried out in \cite{our3,our2011,our2013,indenerN,GorkaKsi}, while the corresponding studies of the vacuum magnetic flux can be found in \cite{Gorka16,Gorka22,Gorka22PRD}.

In the case of a spinor matter field  (which is multi-component, unlike a scalar field), the above-mentioned intuitively understandable boundary conditions cannot be applied, as they lead to mathematical inconsistencies. To achieve mathematical consistency of the theory, we need to use the condition of self-adjointness for the Dirac Hamiltonian operator. This condition will automatically guarantee the impermeability of the tube to matter fields.

In the case of $(2+1)$-dimensional space-time, this requirement leads to a one-parameter boundary condition, see \cite{SitGor19}. But in the physically interesting case of $(3+1)$-dimensional space-time, it leads to a four-parameter boundary condition, see \cite{Akhmerov,Hashimi,Sit2015}. Moreover, the parameter values can, in general, vary along different points of the linear string. However, imposing invariance requirements on discrete symmetries reduces the number of boundary condition parameters \cite{Sit2021flux}.

We will be interested in the induction of magnetic flux in the vacuum of a matter field. The relevance of this problem is related to the fact that the existence of magnetic cosmic strings is one of the possible theories \cite{Horiguchi,Subr} that could solve the problem of the existence of magnetic fields in intergalactic voids \cite{Nero}. Since the induction of magnetic flux in the vacuum of a scalar field was considered in detail for the case of bosonic matter \cite{Gorka22PRD}, in this paper, we will be interested in the case of spinor matter.

In this paper, we will extend the results of \cite{Sit2021flux}. Namely, we will consider the induced magnetic field flux in the vacuum of a spinor massive field in $(3+1)$-dimensional space-time in the background of a cosmic string modeled as impenetrable for the matter field tube of finite radius $r_0$. Taking into account the small value of the cosmic string tension (at least $G\mu \lesssim 10^{-6}$), we shall analyze vacuum polarization in flat Minkowski space-time, while treating the parameter ${e}\Phi/(2\pi)=n e/e_H$ as arbitrary. We will construct graphical representations of the induced magnetic flux and carefully analyze the choice of parameters that define the boundary condition on the surface of the magnetic tube.

\section{Induced magnetic flux}

The operator of the second-quantized spinor field in flat space-time outside the linear topological defect has the form
\begin{equation}\label{1.1}
\Psi(x^0,\textbf{x})=\sum\hspace{-1.4em}\int\limits_{\lambda}\frac1{\sqrt{2E_{\lambda}}}\left[e^{-iE_{\lambda}x^0}\psi_{\lambda}(\textbf{x})\,a_{\lambda}\!+\!
  e^{iE_{\lambda}x^0}\psi_{-\lambda}(\textbf{x})\,b^\dag_{\lambda}\right],
\end{equation}
where $a^\dag_\lambda$ and $a_\lambda$ ($b^\dag_\lambda$ and $b_\lambda$) denote the creation and annihilation operators for spinor particles (antiparticles), which satisfy the standard anticommutation relations;  $\lambda$ is the set
  of parameters (quantum numbers) specifying the state;  $E_\lambda=E_{-\lambda}>0$ is the energy of the state; $\psi_\lambda({\bf x})$ represents a solution of the stationary Dirac equation
\begin{equation}\label{1.5}
 H \psi_{\lambda}({\bf x})=E_\lambda \psi_{\lambda}({\bf x}),
\end{equation}
and symbol \mbox{$\displaystyle \sum\hspace{-1.4em}\int $} denotes a combined operation: summation over the discrete part and integration (with a certain measure) over the continuous values of $\lambda$.

In the case of the cosmic string background and flat Minkowski space-time, the Dirac Hamiltonian takes the form
\begin{equation}\label{1.6}
H=-{\rm i} \gamma^0 \mbox{\boldmath
$\gamma$}\cdot \left(\mbox{\boldmath
$\partial$} - {\rm i}\tilde e\, \textbf{V}\right)+\gamma^0 m,
\end{equation}
where $\tilde e$ is the coupling of the spinor field to the gauge field $U_X(1)$.
In cylindrical coordinates $(r, \varphi, z)$, with the symmetry axis aligned with that of a straight cosmic string, only the angular component of the field potential is non-vanishing
\begin{equation}\label{2.10}
    V_\varphi=\frac{\Phi}{2\pi}, 
\end{equation}
and Hamiltonian \eqref{1.6} obtains form
\begin{equation}\label{2.11}
H=-{\rm i}\gamma^0
\left[\gamma^r \partial_r +\gamma^\varphi\left(\partial_\varphi-{\rm
i}\frac{\tilde e \Phi}{2\pi}\right)+\gamma^3 \, \partial_z\right]+\gamma^0 m.
\end{equation}

This Hamiltonian can be presented as \cite{Sit2021flux} 
\begin{equation}\label{2.15}
H=\left(\begin{array}{cc}
                    H_1 & - {\rm i}\sigma^1 \partial_z\\
                    - {\rm i}\sigma^1 \partial_z & H_{-1}
                     \end{array}\right),
\end{equation}
where
\begin{equation}\label{2.16}
H_s=-{\rm i}\left[\left(s \sigma^1\sin\varphi - \sigma^2\cos\varphi\right)\partial_r \vphantom{\frac12} + \frac{1}{r}\left(s \sigma^1\cos\varphi  + \sigma^2\sin\varphi\right)\left(\partial_\varphi-{\rm
i}\frac{\tilde e \Phi}{2\pi}\right)\right]+\sigma^3 m
\end{equation}
and parameter $s= \pm 1$. The four-component function, $\psi_E(r,\varphi, z)$, can be decomposed  into the two two-component functions, $\psi_E^{(1)}(r,\varphi)$  and $\psi_E^{(-1)}(r,\varphi)$,
\begin{equation}\label{2.17}
\psi_E(r,\varphi, z) =\frac{e^{{\rm i}k_3 z}}{\sqrt{2\pi}}                    \left(\begin{array}{c}
                   \psi_E^{(1)}(r,\varphi) \\
                   {\rm i} \psi_E^{(-1)}(r,\varphi)
                    \end{array}\right)
\end{equation}
and
\begin{equation}\label{2.24}
\psi_E^{(s)}(r,\varphi) = \sum_{n \in \mathbb{Z}}
                   \left(\begin{array}{c}
                   f_n^{(s)}(r,E ) \, \exp\left[{\rm i}\left(n + \frac12 - \frac12 \, s\right)\varphi\right] \\
                   g_n^{(s)}(r,E ) \, \exp\left[{\rm i}\left(n + \frac12 + \frac12 \, s\right)\varphi\right]
                    \end{array}\right),
\end{equation}
where $\mathbb{Z}$ is the set of integer numbers.

The induced vacuum current density is given then as
\begin{equation}\label{1.7}
\textbf{j}(\textbf{x})=\langle {\rm vac}| \Psi^\dag(\textbf{x},t)
\gamma^0 \mbox{\boldmath
$\gamma$}   \Psi(\textbf{x},t) |{\rm vac}
\rangle=-\frac12\sum\hspace{-1.4em}\int \rm{sgn}(E)
\psi^\dag_E(\textbf{x}) \gamma^0 \mbox{\boldmath
$\gamma$}   \psi_E(\textbf{x}).
\end{equation}
It can be shown that $j_r =0$ and $j_z =0$, and
\begin{equation}\label{1.7}
\!j_\varphi\!=\!
- r\! \sum_{s = \pm 1}\!\sum\hspace{-1.4em}\int \sum_{n \in
\mathbb{Z}}\! {\rm sgn} (E)  s  f_n^{(s)}(r,E) \, g_n^{(s)}(r,E).
\end{equation}

It should be noted that function \eqref{2.24} is similar to the spinor function in the case of $(2+1)$-dimensional space-time, where parameter $s$ can be $1$ or $-1$. The induced vacuum current density in \eqref{1.7} is also presented as the sum of the induced vacuum currents in the case of $(2+1)$-dimensional space-time \cite{SitGor19} and additional integration over the momentum component directed along the string.

The magnetic field strength is induced in the vacuum due to the Maxwell equation
\begin{equation}\label{1.8}
\mbox{\boldmath $\partial$}\times \textbf{B}_{\rm I}(\textbf{x}) =
e\, \textbf{j}(\textbf{x}),
\end{equation}
where $e$ is the electromagnetic coupling constant. So, we get
\begin{equation}\label{2.30}
B_{\rm I}(r) = e \int\limits_r^\infty \frac{dr'}{r'} \,
j_\varphi(r')
\end{equation}
and the total induced vacuum magnetic flux is
\begin{equation}\label{2.31}
\Phi_{\rm I} =  \int\limits_{0}^{2\pi} d \varphi \int\limits_{r_0}^\infty dr\, r B_{\rm
I}(r).
\end{equation}

\section{Boundary conditions}

As it was shown in \cite{Sit2015,Sit2021flux}, in the case of the straight cosmic string of radius $r_0$ in $(3+1)$-dimensional space-time, the boundary condition on the tube's surface depends on four parameters $(u,v,t^z,t_\varphi)$ and has form\vspace{-0.5em}
\begin{equation}\label{3.4}
(I-K)\left.\psi \right|_{r = r_0} = 0,
\end{equation}
where
\begin{multline}\label{3.5}
K= \frac12\left[u^2-v^2-(t_\varphi)^2/r^2-(t^z)^2\right]^{-1/2}\times\\ \left(\left[1+u^2-v^2-(t_\varphi)^2/r^2-(t^z)^2\right]I+\left[1-u^2+v^2+(t_\varphi)^2/r^2+(t^z)^2\right]\gamma^0\right)  \\
\times\left({\rm i}u\gamma^r-{\rm i}v\gamma^5-t_\varphi\gamma^\varphi-t^z\gamma^z\right).
\end{multline}
This condition guarantees the self-adjointness of the Hamiltonian operator for the matter field and enforces the impenetrability of the tube. It is worth emphasizing that the parameters defining this boundary condition may vary along the length of the string.

Imposing requirements for invariance with respect to discrete symmetry ($P$, $C$, $CT$, $CPT$) transformations reduces the number of boundary condition parameters \cite{Sit2021flux}. But using only the requirements of $P$ and $CT$ invariance leads to an infinite value of the induced vacuum charge near the string, which is physically undesirable. 
The requirement of $CPT$ invariance guarantees that both the induced vacuum charge density and the total induced vacuum charge along the string vanish, while the induced vacuum magnetic flux remains infinite. In this case, we can write the boundary condition as only one-parameter relation between components of spinor functions \eqref{2.24}
\begin{equation}\label{3.19}
\left.f_n^{(s)}\right|_{r=r_0}+\tan\left(s\frac{\theta}{2}+\frac{\pi}{4}\right)\left.g_n^{(s)}\right|_{r=r_0}=0,
\end{equation}
where $u=\sec\theta$ and $t^z=t_\phi=v=0$. 

Finally, the requirement of the charge conjugation ($C$) invariance ensures that the induced charge along the string vanishes and the induced vacuum magnetic flux remains finite. 
As it was pointed out in \cite{Sit2021flux}, this requirement restricts the boundary parameter $\theta$ to only two possible values: $\theta = 0$ and $\theta = \pi$. An important point is that relation \eqref{3.19} is the same as in the case of $(2+1)$-dimensional space-time for $s=1$. As shown in \cite{SitGor19}, in this low-dimensional case, the induced vacuum flux is finite only at $\theta=0$ and $\theta=\pi$ too.

\section{Induced magnetic flux}

In order to analyse the induced magnetic flux in the vacuum of the spinor field in the cosmic string background in $(3+1)$-dimensional space-time, we will use the analytical result of \cite{Sit2021flux} for the cases $\theta = 0$, $\theta = \pi$, and \vspace{-0.5em}
\begin{equation}\label{labelF}
     F = \left\{\!\!\left| \frac{\tilde
e \Phi}{2\pi}  \right|\!\!\right\}\neq \frac12.
\end{equation}
In this case, it was obtained \vspace{-0.5em}
\begin{equation}\label{5.16}
\Phi_{\rm I}(mr_0)= \int\limits_0^\infty\,\frac{dk_3}{\pi}\, \Phi^{(2dim)}_{\rm I}\!\!\left(\!\sqrt{m^2+k_3^2}\,r_0\right), 
\end{equation}
where $k_3$ is the momentum component
directed along the string, $r_0$ is the radius of the tube,  $\Phi^{(2dim)}_{\rm I}(mr_0)$ is the induced magnetic flux in the case of $(2+1)$-dimensional space-time that was obtained in \cite{SitGor19}, see Exp.(5.17) there at $\nu=1$. For the case of $F=1/2$ and $(3+1)$-dimensional space-time it was shown that only at $\theta = 0$ the induced magnetic flux is finite and equal to zero.

It is worth noting that $\Phi^{(2\mathrm{dim})}_{\rm I}(mr_0)$ decreases as $m r_0$ increases for the boundary parameter $\theta=0$, while it increases for $\theta=\pi$~\cite{SitGor19}. In $(2+1)$-dimensional space-time, both cases are physically admissible, as they yield finite values of the induced vacuum magnetic flux. But in the case of $(3+1)$-dimensional space-time, the boundary parameter $\theta=\pi$ is not acceptable, as it inevitably leads to an infinite value of the induced magnetic flux. This divergence arises from the integral of the increasing function in \eqref{5.16}, see Fig.2 in \cite{SitGor19}. The conclusion in \cite{Sit2021flux} that the boundary condition parameter $\theta=\pi$ is admissible in $(3+1)$-dimensional space-time, and that the induced magnetic flux in this case grows with increasing tube thickness, was made in error, since the flux diverges for tubes of arbitrary thickness. 

Thus, in the following, we will only consider the case of the boundary condition parameter $\theta=0$. 
In this case, the boundary condition on the tube's surface takes the form
\begin{equation}\label{3.4theta0}
(I-{\rm i} \gamma^r)\left.\psi \right|_{r = r_0} = 0,
\end{equation}
or as relation between components of spinor functions
\begin{equation}\label{3.19theta}
\left.f_n^{(s)}\right|_{r=r_0}+\left.g_n^{(s)}\right|_{r=r_0}=0.
\end{equation}
Note that the four parameters of the boundary condition of general form \eqref{3.5} in this case are fixed along the tube's surface and written as $t^z=0$, $t_\phi=0$, $v=0$, $u=1$.

\begin{figure}[t]\centering
	\includegraphics[width=\textwidth]{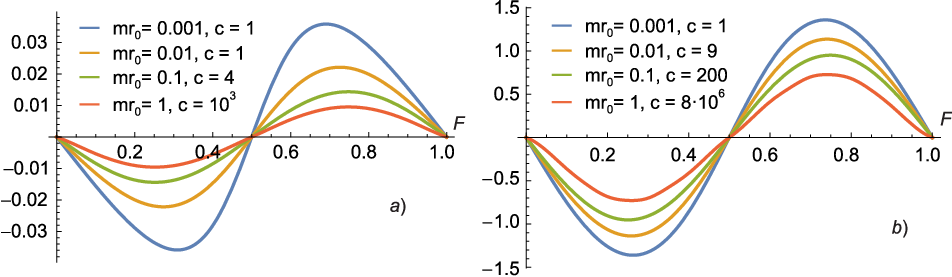}
    \caption{Induced vacuum magnetic flux in dimensionless units as the function of the "magnetic" flux inside the tube $F$ at boundary parameter $\theta=0$ for different values of the tube's thickness $mr_0=10^{-3}$, $10^{-2}$, $10^{-1}$, $1$: \textit{a}) $ e m^{-1}\Phi^{(d=2)}_{\rm I}$ in $(2+1)$-dimensional space-time, \textit{b}) $ e \pi \Phi^{(d=3)}_{\rm I}$ in  $(3+1)$-dimensional space-time.  For convenience of presentation, the above functions are multiplied by the coefficient $c$.}
	\label{Fig1}
\end{figure}

We consider the dependence of the induced magnetic flux on flux inside the tube \eqref{labelF} for the case of $(3+1)$-dimensional space-time and compare it with the case of $(2+1)$-dimensional space-time. The results of our computations are presented in Fig.\ref{Fig1}.

As can be seen, the presented functions are symmetric under the replacement $F \longleftrightarrow 1-F$. They vanish at $F=0$, $1/2$, $1$, while their extremum values shift slightly toward $F=1/2$ as the tube thickness decreases. 

We also consider the dependence of the induced magnetic flux on the tube's radius at fixed "magnetic" flux inside the tube \eqref{labelF}  $F=0.7$ for the case of $(3+1)$-dimensional space-time and compare it with the case of $(2+1)$-dimensional space-time, see Fig.\ref{pic1}. The choice of the value of $F$ is due to the fact that the value of the induced magnetic flux at $F=0.7$ is close to the maximum for the tube of different thickness.  As one can see, for the small tube radius $mr_0 \leq 0.16$, the induced dimensionless magnetic flux in $(3+1)$-dimensional space-time is greater than in the case of $(2+1)$-dimensional space-time.

\section{Conclusions}

In this paper, we considered the effect of inducing magnetic flux in the vacuum of a charged massive spinor field in the case of $3+1$-dimensional space-time in the background of a magnetic cosmic string based on the results \cite{SitGor19,Sit2021flux}. The cosmic string was modeled as an impenetrable tube of finite thickness for the matter field, containing a ``magnetic'' field inside and subject to a certain boundary condition on its surface. The gauge field inside the tube can be some Abelian field of the early Universe. We demonstrated that in $(3+1)$-dimensional space-time, irrespective of the internal structure of the cosmic string, the only admissible boundary condition on the tube's surface corresponds to the parameter $\theta=0$ in \eqref{3.19}. This corresponds to the
MIT quark bag boundary condition \cite{Chodos,Johnson}. At all other boundary conditions, the induced in spinor vacuum charge or magnetic flux will be infinite. The statement made in \cite{Sit2021flux} that the value of the boundary condition parameter $\theta=\pi$ is also physically permissible, and that the induced vacuum magnetic flux in this case grows with increasing tube thickness, turned out to be false, since the induced flux diverges for tubes of arbitrary thickness.

It should be noted that in the case of $(2+1)$-dimensional space-time, two possible values of the boundary condition parameter are allowed:  $\theta=0$ and $\theta=\pi$. 
In the case of $\theta=0$, the induced magnetic flux decreases with increasing tube thickness. In contrast, for $\theta=\pi$, the induced magnetic flux remains finite but increases with tube thickness, which appears somewhat strange from a general physical perspective.

\begin{figure}[t]\centering
	\includegraphics[width=\textwidth]{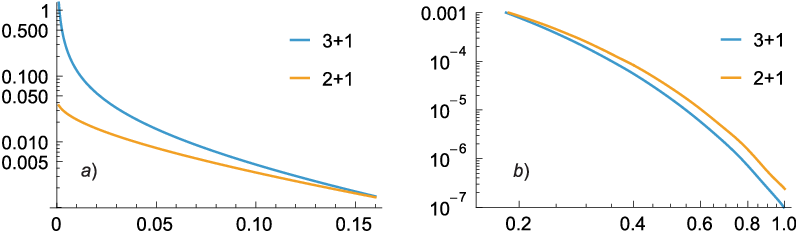}
    \caption{Induced vacuum magnetic flux in dimensionless units in 2+1 space-time ($ e m^{-1}\Phi^{(d=2)}_{\rm I}$), and in 3+1 space-time $ e \pi \Phi^{(d=3)}_{\rm I}$ as function of the tube thickness $mr_0$ for the cases of $mr_0<0.16$ (\textit{a}) and $mr_0>0.16$ (\textit{b}).}
	\label{pic1}
\end{figure}

Results of our numerical computations for the case of $(3+1)$-dimensional space-time are presented in Fig.\ref{Fig1} and Fig.\ref{pic1}, and compared with the case of $(2+1)$-dimensional space-time. In Fig.\ref{Fig1}, the sine-like dependence of the induced in spinor vacuum magnetic flux on the "magnetic" flux inside the tube coincides with the sine-like dependence for the case of $(2+1)$-dimensional space-time \cite{SitGor19}, but has a greater amplitude for the small tube thickness $mr_0\leq 0.16$. The strongly increasing behavior of the dimensionless induced in vacuum magnetic flux for the case of $(3+1)$-dimensional space-time compared to the case of $(2+1)$-dimensional space-time at small thickness of the tube $mr_0\ll 1$, see Fig.\ref{pic1}, can be explained in the approach of zero thickness magnetic tube (singular magnetic vortex), see, e.g., \cite{Sit96} for spinor and \cite{SitVlas} for scalar matter. As one can see, for the case of $(2+1)$-dimensional space-time the induced vacuum magnetic flux tends to a constant at $mr_0\rightarrow 0$ \cite{Sit96,SitVlas}, but for the case of $(3+1)$-dimensional space-time induced vacuum magnetic flux is divergent $\sim \ln(mr_0)$ at $mr_0\rightarrow 0$ \cite{SitVlas}.

Finally, we note that the induction of magnetic flux in the vacuum of a spinor field (with mass of the spinor particle $m$) will be significant only if \mbox{$mr_0\!\ll\! 1$.} The thickness of the cosmic string can be found from the approximate condition $m_H r_0= 1$, where $m_H$ is the scale of spontaneous symmetry breaking (or mass of the corresponding Higgs field) producing a linear topological defect (cosmic string). So, we can write the relation $mr_0=m/m_H$. It will be less than 1 if $m \ll m_H$, which is plausible for fermion fields of the Standard Model. Moreover, the fields of light fermions will have a stronger effect on vacuum polarization in the background of magnetic cosmic strings.

\vspace{5mm}

\centerline{\bf Acknowledgements}
\vspace{5mm}

The work of V.M.G was partially supported by the project 'High-energy processes in plasma: acceleration of cosmic rays and their contribution to space weather' of the Ministry of Education and Science of Ukraine (25BF051-04). The authors are grateful to T.V. Gorkavenko and M.S. Tsarenkova for helpful comments and help in the preparation of this paper.

\bibliographystyle{JHEP}
\bibliography{main.bib}

\end{document}